\documentclass[aps,pre,longbibliography,notitlepage,floatfix,11pt,tightenlines,nofootinbib]{revtex4-1}

\usepackage{enumitem}
\usepackage{bm,amsmath,amssymb,graphicx,stmaryrd,subcaption}
\usepackage{tikz}
\usepackage[abs]{overpic}
\newcommand{\uvc}[1]{\bm{\mathrm{\hat #1}}} 

\DeclareMathOperator{\sech}{sech}
\DeclareMathOperator\arctanh{arctanh}
\DeclareMathOperator\arcsinh{arcsinh}

\usepackage[format=plain,labelfont={bf,small},textfont=small,justification=raggedright,singlelinecheck=false]{caption}
\begin{document}

\title{Anisotropic swelling of anisotropic elastic panels}
\author{H. G. Wood}
\affiliation{Moog Inc., 1213 North Main St., Blacksburg, VA 24060}
\author{J. A. Hanna}
\email{jhanna@unr.edu}
\affiliation{Department of Mechanical Engineering, University of Nevada, 1664 N. Virginia St.\ (0312), Reno, NV 89557-0312}

\date{\today}

\begin{abstract}
While isotropic in-plane swelling problems for thin elastic sheets have been studied extensively in recent years, many shape-programmable materials, including nematic solids and 3D-printed structures, are anisotropic, as are most industrial sheet materials.
In this theoretical work, we consider central swelling and shrinkage of plates of aspect ratio and material properties relevant to the manufacture of engineered wood composite panels in which both in-plane swelling and material stiffness are highly orthotropic, leading to multiple separations in energy scales. 
With transverse swelling in the soft direction, and gradients in the stiff direction, the warped plates adopt two distinct types of configurations, axisymmetric and twisted, which we illustrate with toy models.
We employ a two-parameter family of isometries to embed the metric programmed by the swelling, thus reducing the problem to one of minimizing bending energy alone.
A simple argument is seen to closely predict averaged axisymmetric curvatures.
While purely cylindrical shapes are unobtainable by pure in-plane swelling, they can be closely approximated in a highly anisotropic system.
However, anisotropy can favor twisting, and breaks a degenerate soft deformation mode associated with minimal surfaces in isotropic materials.
Bifurcations from axisymmetric to twisted shapes can be induced by anisotropy or by certain attributes of a central shrinkage profile.
Finally, we note how our findings indicate practical limitations on the diagnosis of moisture inhomogeneities in manufactured panels by observation of warped conformations, due to the sensitivity of the qualitative response to specifics of the profile.
\end{abstract}
\maketitle

\section{Introduction}

Many natural and industrial soft materials are anisotropic both in their elastic stiffness and their propensity to swell in response to moisture or other stimuli.
However, despite much recent interest in the programmed shaping of thin sheets by swelling, relatively little is known about the effects of anisotropy on such approaches.
Studies exist on transitions to twisted shapes in thin nematic elastomer strips with through-thickness gradients in orientation \cite{TeresiVarano13, TomassettiVarano17, Sawa11}, the
behavior and shape-programmability of bilayers of anisotropic materials \cite{Nardinocchi15, Battista19-1, VanRees17}, and the effect of the aspect ratio of a consistently oriented perimeter on the shapes adopted by bilayers of pre-stressed isotropic materials \cite{Caruso18}.
The warpage and bi-stability of carbon-fiber composite laminates are well known phenomena \cite{Hyer81-2}.  
Moisture-induced warpage of natural and engineered wood \cite{woodhandbook, finnishhandbook} panels is of interest both as a problem in manufacturing \cite{Rindler17, Boulton20, Brouse61, Heebink64, Grossman73, SuchslandMcNatt86, FridleyTang92, Lang95, Ormarsson98, SmithGramberg02, Ganev03, CaiDickens04, Gereke09} and as a tool for the design of adaptive architecture \cite{Correa15, Abdelmohsen19}.  

The current study differs from the above works in several ways.  
We examine a basic problem in anisotropic swelling, that nonetheless reveals new phenomena.
Our system is a thin plate composed of a single layer of highly anisotropic material, with a consistent shape and orientation of its perimeter and no through-thickness gradients in either properties or swelling.  
Among a class of axisymmetric and twisted mid-surface embeddings and simple swelling distributions, 
anisotropy is seen to influence the shape of the body, alongside other factors.
While inspired and guided by a specific industrial problem, the findings are generally applicable to any thin bodies in which one direction is much stiffer than the other in terms of both bending and response to swelling stimuli.

\section{Overview of the problem}

\begin{figure}[h]
	\centering
	\includegraphics[width=0.45\linewidth]{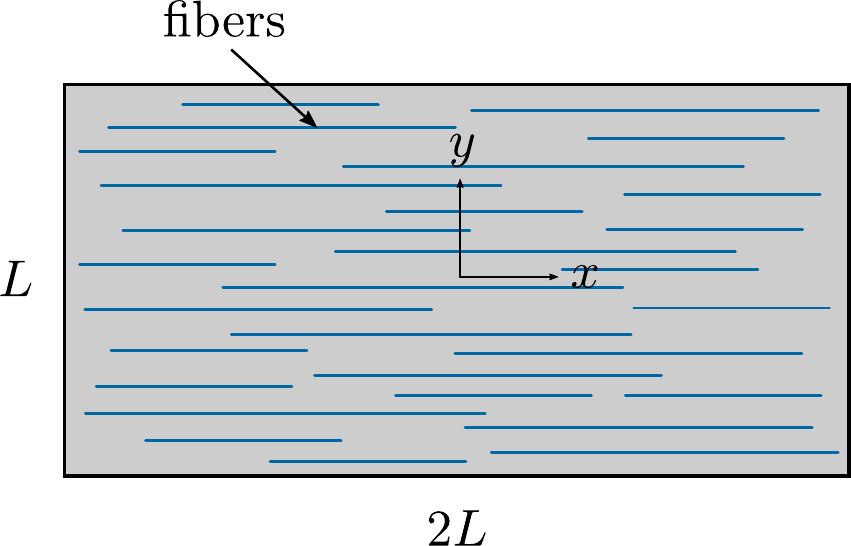}
	\caption{Schematic of an unwarped 2:1 rectangular panel with fibers oriented along the long axis.  The coordinates $x$ and $y$ are attached to the material, with origin in the middle of the plate, and oriented in the long stiff and short soft directions, respectively.  The panel is homogeneous in the through-thickness direction (here, into the page).   }
	\label{anplate}
\end{figure}

The parameters of our problem are chosen to mimic the geometry and material properties of a typical thin rectangular panel of laminated veneer lumber (LVL), a wood composite formed from several layers of identically oriented orthotropic material.  Figure \ref{anplate} is a schematic of such a body, with a length $2L$ in the stiff fiber ($x$) direction and a width $L$ in the soft, transverse ($y$) direction.  
We treat this as an effectively two-dimensional body, with no variation in properties or swelling in the third, through-thickness, direction.
The coordinates $x$ and $y$ have their origin in the middle of the plate, and are attached to the material so that they convect with it as the panel swells-- they are not Cartesian coordinates.  Swelling, whether positive or negative (shrinkage), is entirely in the soft direction.  This approximation is inspired by the fact that swelling along fiber directions in wood is one to two orders of magnitude smaller than swelling in other directions \cite{Peck57}.

A simple question to ask about this system is, given a patch of swelling or shrinkage in the center of the panel, what curved shape will the panel adopt?
It is well known that for an isotropically swelling isotropically elastic material, central swelling and shrinkage will lead to dome-like and saddle-like configurations, respectively.   It is easy to imagine variants of such motifs, with curvatures biased by anisotropy in bending stiffness.  In addition, anisotropic plates admit a qualitatively different form in which stiff fibers twist around a soft direction.  These general classes of shapes will be discussed further in Section \ref{results}.

For analytical convenience, we consider swelling in the soft $y$ direction with swelling gradients in the stiff $x$ direction.
We represent this by saying that the mid-surface rest or target metric \cite{Efrati09} has the line element $dx^2 + U(x)^2dy^2$.  This also implies that $x$ is the arc length along coordinate lines in the fiber direction.
The swelling function $U(x)^2$ is chosen to have a central peak or valley and be symmetric in $x$.

\section{Energy}

Following standard practice, we presume that the energy for a plate of thickness $h \ll L$ takes the form of an integral over the mid-surface,  
\begin{align}
	E = \int \! dA \, \frac{1}{2}C^{\alpha\beta\gamma\delta}\left(h\varepsilon_{\alpha\beta}\varepsilon_{\gamma\delta} + \frac{h^3}{12}b_{\alpha\beta}b_{\gamma\delta}\right) \, ,
	\label{energy}
\end{align}
where $\varepsilon_{\alpha\beta}$ and $b_{\alpha\beta}$ are components of mid-surface stretching and bending tensors, and the indices run from 1 to 2.
In general, the forms of these tensors depend on the constitutive model \cite{WoodHanna19}, but this issue will not arise in the present work, as we will require the mid-surface strains to vanish.  
A typical engineered wood panel is $8$x$4$ feet and on the order of an inch thick, that is, one to two orders of magnitude thinner than its lateral dimensions.
For such a panel, there is a primary separation of scales between stretching and bending energies, and we can eliminate the former entirely by choosing only surfaces that embed the swollen target metric.
Thus, the $\varepsilon_{\alpha\beta}$ all vanish identically, the area form $dA = Udxdy$, and in the remaining bending component of the energy the $b_{\alpha\beta}$ will be components of curvature.  As the panel is plate-like, with no through-thickness variations, its rest curvature is everywhere zero.
For the elastic tensor, we covariantize an established form for the elastic tensor of a five-modulus transversely isotropic material \cite{Spencer84} using the inverse metric components $g^{\alpha\beta}$ of the swollen target metric and the contravariant components $q^\alpha$ of a unit vector in the fiber direction, 
\begin{align}
		C^{\alpha\beta\gamma\delta} &=  \lambda g^{\alpha\beta}g^{\gamma\delta} + \mu_T\left(g^{\alpha\gamma}g^{\beta\delta} + g^{\beta\gamma}g^{\alpha\delta}\right) + A \left(q^\gamma q^\delta g^{\alpha\beta} + q^\alpha q^\beta g^{\gamma\delta}\right) \nonumber \\
		&+ \left(\mu_L - \mu_T\right)\left(q^\alpha q^\gamma g^{\beta\delta} + q^\alpha q^\delta g^{\beta\gamma} + q^\beta q^\gamma g^{\alpha\delta} + q^\beta q^\delta g^{\alpha\gamma}\right) + B q^\alpha q^\beta q^\gamma q^\delta \, .
\end{align}
Thus, we neglect any changes in the moduli caused by swelling, an approach that goes in hand with our integration of the energy over the swollen area form; a more precise treatment should likely rethink these assumptions, at least for large swelling amplitudes. 
The resulting energy can be further simplified by noting that $q^x = q_x = 1$ and $q^y = q_y = 0$, and defining  \cite{Hyer81-2} the ``reduced stiffnesses'' $C_{(11)} \equiv \lambda+2\mu_T+2A+4(\mu_L-\mu_T)+B$, $C_{(22)} \equiv \lambda+2\mu_T$, $C_{(12)} \equiv \lambda+A$, and $C_{(66)} = \mu_L$, to obtain
\begin{align}
	E = \int \!\!\! \int \!  Udxdy \, \frac{1}{2}\frac{h^3}{12}\left[C_{(11)}(b_x^x)^2 + 4C_{(66)}b_y^x b_x^y + 2C_{(12)}b_x^x b_y^y + C_{(22)}(b_y^y)^2 \right] \, .
	\label{bendenergy}
\end{align} 
In some materials of interest, $C_{(11)}$ may be significantly larger than the other coefficients, such that a secondary separation of scales is anticipated.
For example, typical values for LVL might provide a factor of 30 between this stiffness coefficient and the next highest one \cite{Janowiak01}.

For a given swelling pattern $U$, we will seek to minimize \eqref{bendenergy} among a class of embeddings to be described below.  The integration over $-L/2 < y < L/2$ will be trivial, leaving a one-dimensional integral over $-L < x < L$ which will only be analytically tractable in special cases.
Finally, we will not concern ourselves with satisfaction of boundary conditions at the edges of the plate.  Whether any related effects would have any significant impact on the energy is unknown, and indeed the presence of small boundary layers in elastic plates is still an open question \cite{Efrati09-2,Guven19}.

\newpage
\section{Embedding}

\begin{figure}[h]
	\centering
	\includegraphics[width=0.9\linewidth]{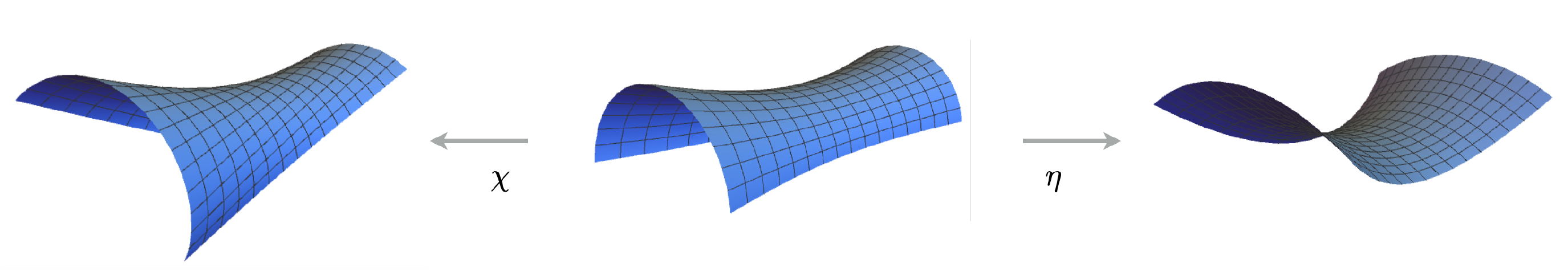}
	\caption{Schematic illustrating the effects of the two parameters on the immersion \eqref{kenmotsu}.  The surface in the center is a piece of an axisymmetric shape.  Changing the parameter $\eta$ changes the ratio of curvatures of this surface.  The parameter $\chi$ twists the surface.}
	\label{embedding}
\end{figure}

Building on an approach used previously in \cite{Kim12softmatter}, we embed the swollen metric by restricting the position $\mathbf{x}$ of the mid-surface of the plate to a two-parameter family of isometric immersions \cite{Kenmotsu03, DoCarmoDajczer82},
\begin{align}
	\mathbf{x}(x,y\,; \eta,\chi) =& \left[\Lambda(x) + \chi\theta(x,y)\right]\uvc{e}_1 + \sqrt{\eta^2 U(x)^2 - \chi^2}\left[\cos\theta(x,y)\,\uvc{e}_2 + \sin\theta(x,y)\,\uvc{e}_3\right] \, ,
	\label{kenmotsu} \\	
	&\;\,\Lambda(x) = \int \! dx \, \frac{\eta U(x)}{\eta^2 U(x)^2 - \chi^2}\sqrt{\eta^2 U(x)^2\left[1 - \eta^2 U'(x)^2\right] - \chi^2} \, , \nonumber \\
	&\;\,\theta(x,y) = \frac{y}{\eta} - \frac{\chi}{\eta}\int \! dx \, \frac{\sqrt{\eta^2 U(x)^2\left[1 - \eta^2 U'(x)^2\right] - \chi^2}}{U(x)\left[\eta^2 U(x)^2 - \chi^2 \right]} \nonumber \, ,
\end{align} 
where the $\uvc{e}_i$ are fixed Cartesian axes, and a prime denotes an $x$-derivative.  We assume $U > 0$ to prevent self-intersections.
Figure \ref{embedding} illustrates the effects of the two parameters. 
The parameter $\chi$ controls the twisting of the surface, and is zero when the surface is axisymmetrically ``rolled up'' around an axis coplanar with any of the stiff fibers.
The parameter $\eta$ controls the extent of the rolling up.  Because an isometry preserves Gaussian curvature, as the structure curls more around the axis it relaxes its curvature along the axis, so in some sense $\eta$ reflects anisotropy in curvature. 
The maximum $\eta$ and $\chi$ are limited in a real embedding, and depend on the form of $U$.

For isotropic elastic bodies with an energy depending on mean curvature alone, the energy of this embedding would increase with twist parameter $\chi$, or remain zero if the family of surfaces is minimal.  In the present case of an anisotropic plate energy, things are qualitatively different, and twist may be favored.

The energy \eqref{bendenergy} of the immersion \eqref{kenmotsu} may be computed using the definitions of the metric components $g_{\alpha\beta} = d_\alpha \mathbf{x} \cdot d_\beta \mathbf{x}$, inverse metric components defined by the relations $g^{\alpha\beta}g_{\beta\gamma} = \delta^\alpha_\gamma$, and mixed curvature components $b^\alpha_\beta = g^{\alpha\gamma}d_\beta d_\gamma \mathbf{x} \cdot\uvc{N}$, where the unit normal $\uvc{N} = d_\alpha \mathbf{x} \times d_\beta \mathbf{x} / \| d_\alpha \mathbf{x} \times d_\beta \mathbf{x} \|$.  One obtains
\begin{align}
		E = \frac{h^3 L}{24} \int_{-L}^L \! dx &\left[C_{(11)}\left(\frac{\left(-\eta^4 U^3 U'' + \chi^2\right)^2}{\eta^4 U^3 \left(\eta^2 U^2\left[1-\eta^2(U')^2\right]-\chi^2\right)}\right) + 4C_{(66)}\left(\frac{\chi^2}{\eta^4U^3}\right) \right. \nonumber \\
		&\left. \;\; - \, 2C_{(12)}\left(U'' -\frac{\chi^2}{\eta^4U^3}\right) + C_{(22)}\left(\frac{1-\eta^2\left(U'\right)^2}{\eta^2U} - \frac{\chi^2}{\eta^4 U^3}\right)\right] \, .
	\label{kenmotsuenergy}
\end{align}
For a given swelling function $U$, the problem is reduced to minimizing \eqref{kenmotsuenergy} with respect to the two adjustable parameters $\eta$ and $\chi$ of the isometry, subject to the limits on these quantities set by the form of the immersion \eqref{kenmotsu}.

\section{Results}\label{results}

The twisting motif is illustrated by a simple, but carefully chosen, example of a swelling function of the form $U(x) = \sqrt{1+(x/l)^2}\,$, $l \ge \eta > 0$ being a parameter; this form also requires $\eta^2 \ge \chi^2$. 
This choice of function yields a metric of the catenoid-helicoid family of minimal surfaces, and the associated energy \eqref{kenmotsuenergy} admits the closed form expression
\begin{align}
		E = \frac{h^3 L}{12} &\left(  \frac{l^2-\eta^2}{l\eta^2} \left[
		C_{(11)} \frac{1}{l}\sqrt{ \frac{ | \eta^4-l^2\chi^2 | }{\eta^2-\chi^2} } \arctanh \left( \frac{L}{l\sqrt{l^2+L^2}} \sqrt{ \frac{ | \eta^4-l^2\chi^2 | }{\eta^2-\chi^2} } \right) + C_{(22)} \arcsinh \frac{L}{l} \right]
		 \right. \nonumber \\
		 &\left. \; +\, \frac{L}{l\sqrt{l^2+L^2}}\left[ \left(C_{(11)} - 2C_{(12)} + C_{(22)}\right)
		- \frac{l^2\chi^2}{\eta^4}\left(C_{(11)} - 2C_{(12)} + C_{(22)} - 4C_{(66)}\right) \right] \right) \, . \label{energyexp}
\end{align}
Informal examination of this function indicates that it is minimized by first letting $\eta$ reach its allowed maximum $\eta = l$, leading to the simplified expression
\begin{align}
	E_{\eta=l} = \frac{h^3 L^2}{12 l \sqrt{l^2+L^2}}\left[\left(C_{(11)}-2C_{(12)}+C_{(22)}\right) - \frac{\chi^2}{l^2}\left(C_{(11)}-2C_{(12)}+C_{(22)}-4C_{(66)}\right)\right] \, , \label{energyexpsimp}
\end{align} 
and then, noting that $C_{(11)}$ is the dominant coefficient, choosing the maximally twisted surface with $\chi = \eta = l$.  Thus, for this choice of swelling function, elastic anisotropy favors a fully twisted shape, a piece of a helicoid. 
If the material were isotropic, the set of coefficients multiplying the twisting parameter $\chi$ would sum to zero, and the twisting isometry would be a soft (zero-stiffness) deformation.
\begin{figure}[h]
	\centering
	\includegraphics[width=0.45\linewidth]{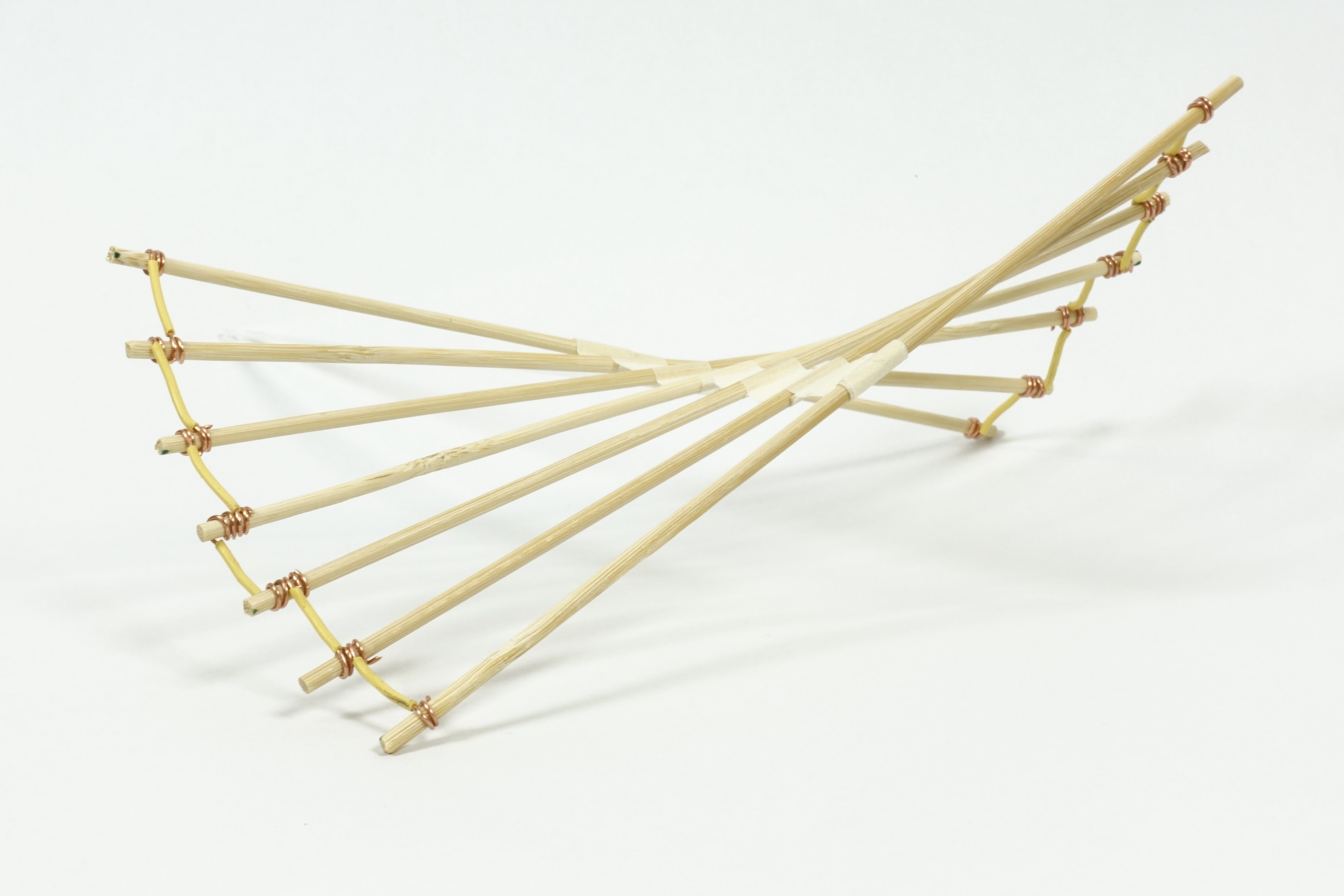}\\
	\includegraphics[width=0.45\linewidth]{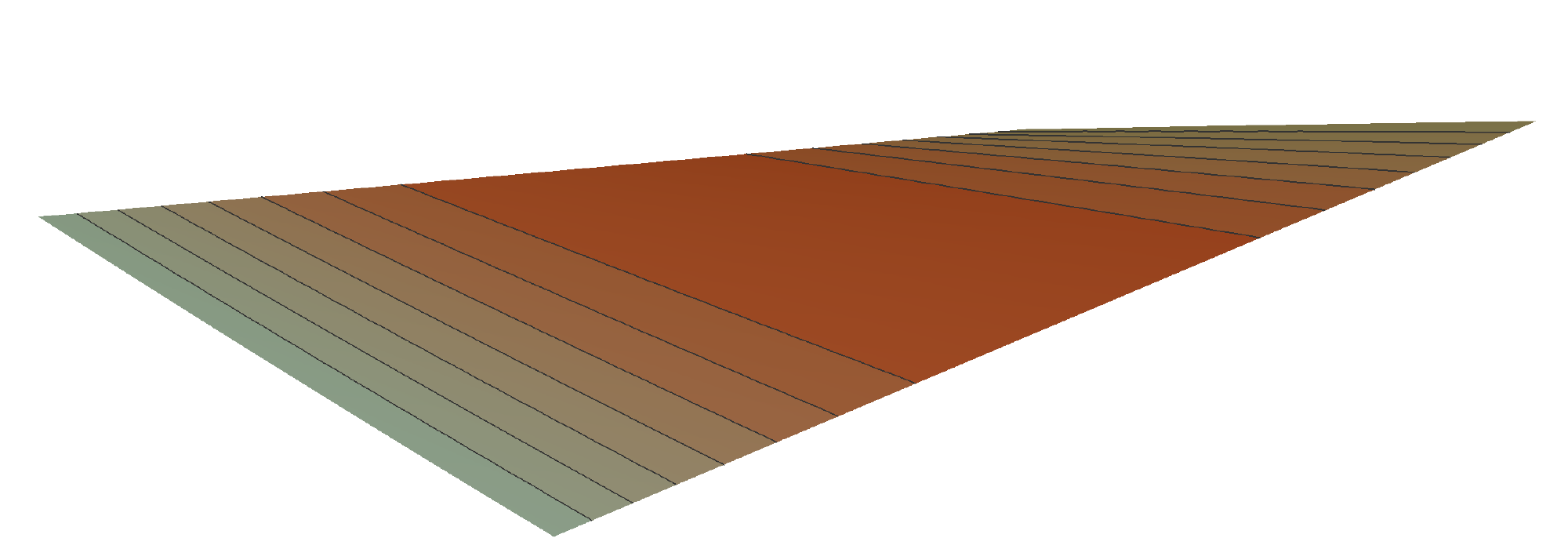}
	\caption{Top: photograph of a toy model of a helicoid-like surface in which stiff directions remain straight while transverse distances increase towards their ends.  Bottom: a minimum energy embedding \eqref{kenmotsu} of the metric with transverse swelling $g_{yy} = U^2 = 1+(x/l)^2$ with parameters $\chi = \eta = l $. 
  Blue indicates more swelling. Lines are contours of $g_{yy}$. }
	\label{twistmodel}
\end{figure}

Twisting allows for a swelling of the surface near its ends while the stiff fiber directions remain straight.  Figure \ref{twistmodel} shows the helicoidal surface alongside an illustrative toy model.  This swelling function can be embedded as a ruled surface, with rulings corresponding to the stiff direction, but this is not possible for general metrics.

\begin{figure}[h!]
	\centering
	\includegraphics[width=0.4\linewidth]{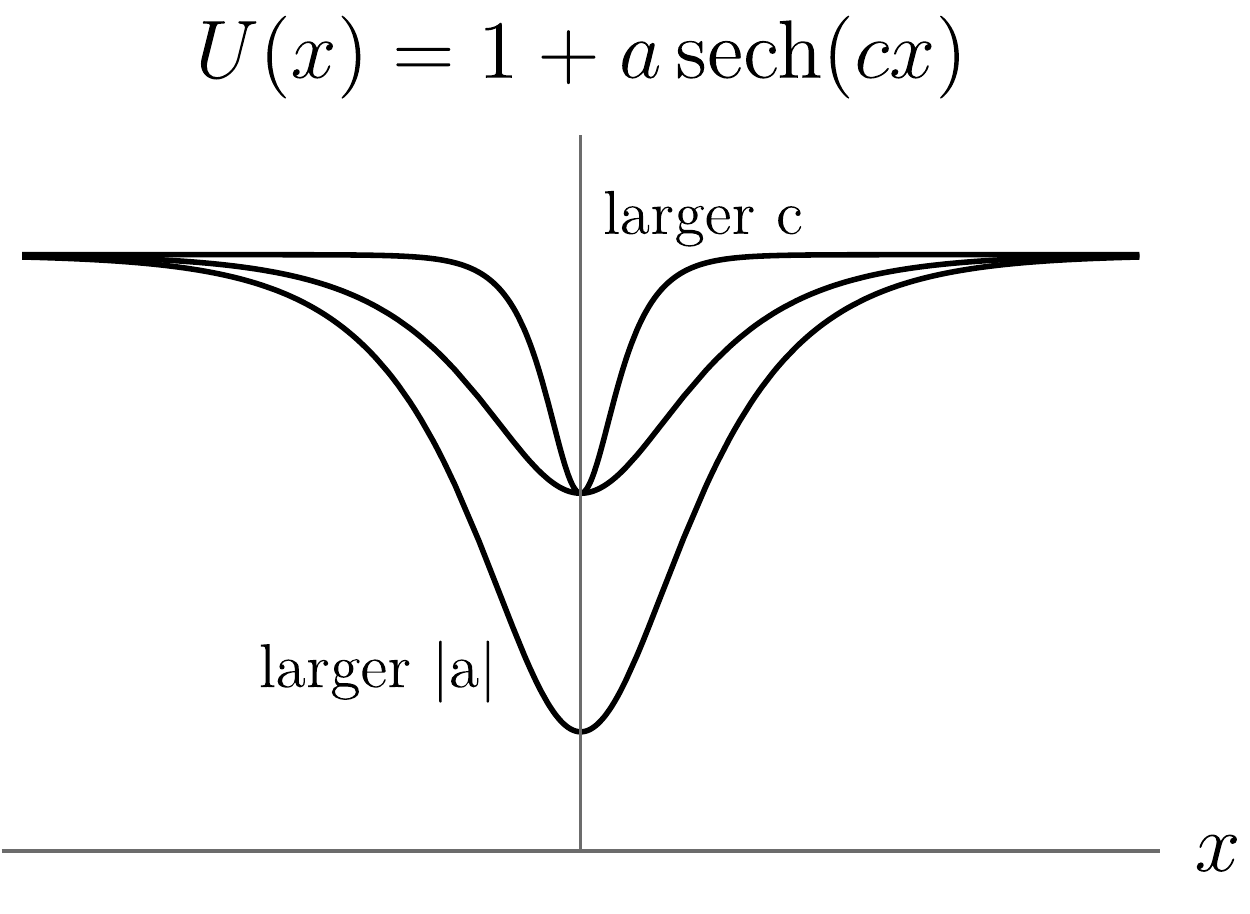}
	\caption{The effect of parameters $a < 0$ and $c$ on the swelling function $U(x)=1+a\sech(cx)$.}\label{bump}
\end{figure}

The competition between axisymmetric and twisted shapes may be examined using a more generic choice of swelling function $U(x)=1+a\sech(cx)$ that allows for either swelling or shrinkage in the middle of the plate, depending on the sign of the amplitude parameter $a$.  The slope parameter $c$ controls the slope of the localized region of central swelling or shrinkage.  The effects of both parameters are illustrated schematically in Figure \ref{bump}.
For the remainder of the paper, we normalize $x$ by $L$ so that $c$ has a clear interpretation.

\begin{figure}[h!]
	\centering
	\includegraphics[width=0.45\linewidth]{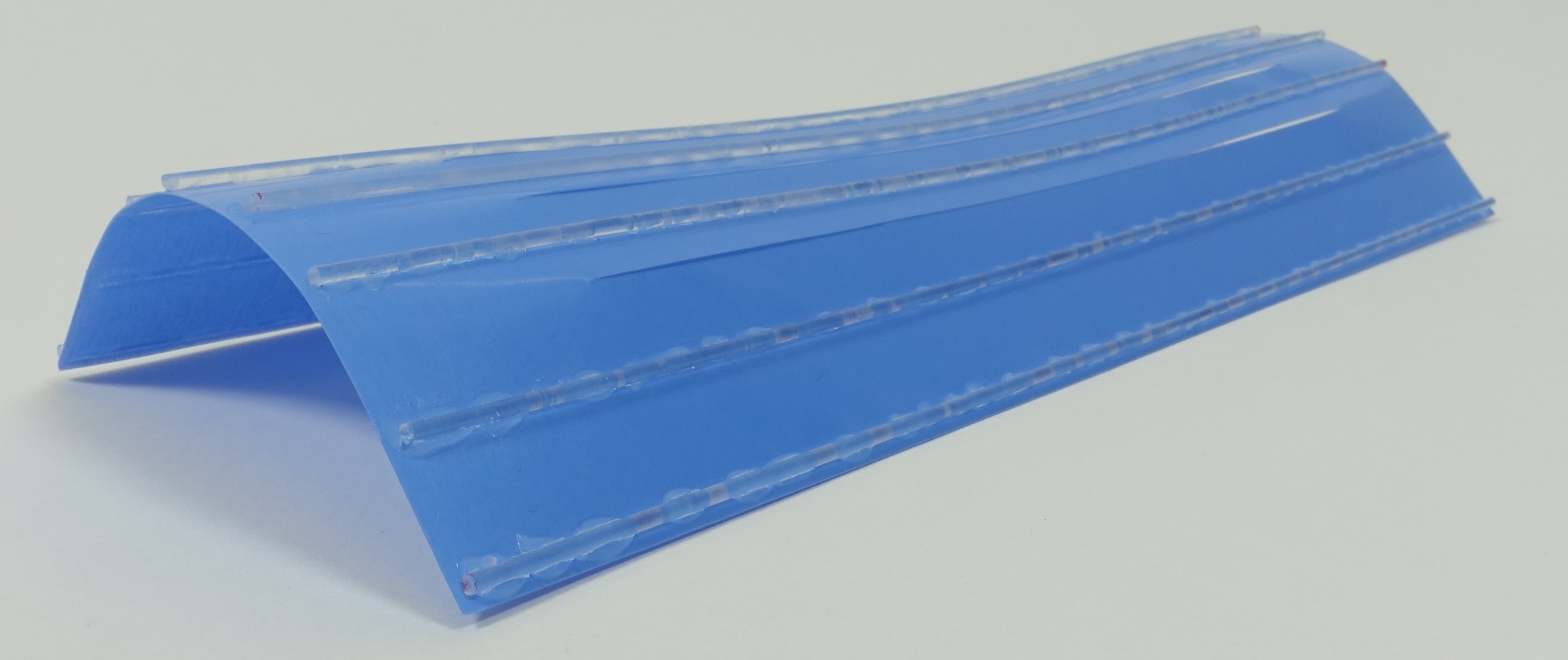}
	\includegraphics[width=0.45\linewidth]{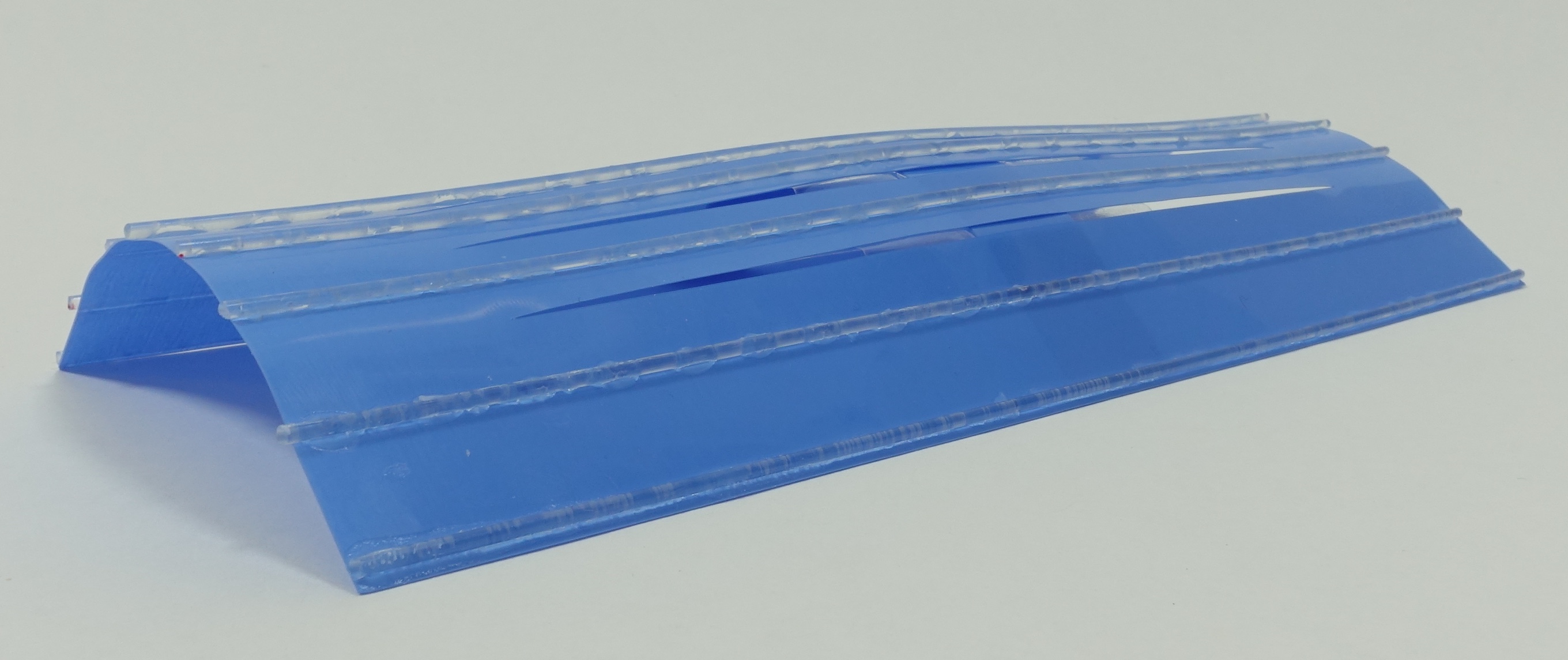}\\
	\vspace{0.1in}
	\includegraphics[width=0.45\linewidth]{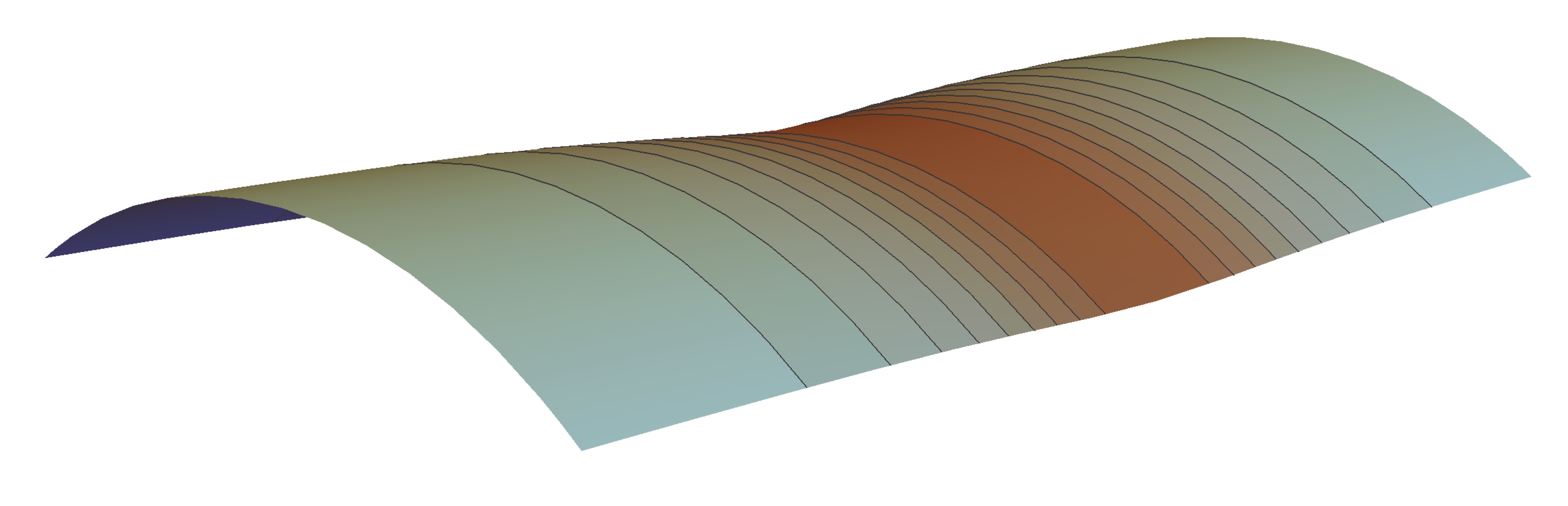}
	\includegraphics[width=0.45\linewidth]{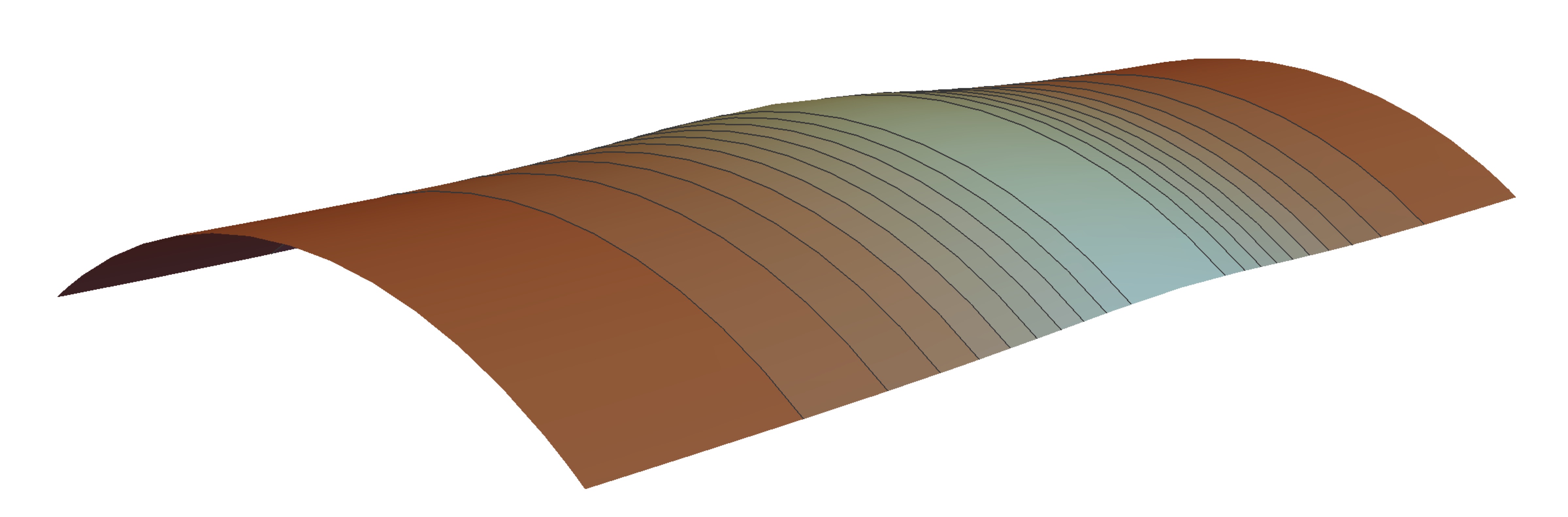}\\
	\caption{ Top: photographs of toy models made from anisotropically stiffened sheets with central contraction (left) or expansion (right).  Bottom: minimum energy embeddings \eqref{kenmotsu} of the metric with transverse swelling $g_{yy} = U^2 = [1+a\sech (cx)]^2$ with parameters $a=-0.05$, $c=4$, $\eta=0.813$, $\chi=0$ (left) and $a=0.05$, $c=4$, $\eta=0.789$, $\chi=0$ (right).  Blue indicates more swelling.  Lines are contours of $g_{yy}$.}
	\label{models}
\end{figure}

Examples of numerically determined minima of the energy \eqref{kenmotsuenergy} are shown in Figure \ref{models}, alongside illustrative toy models.  These axisymmetric ($\chi=0$) shapes result from parameter values $a= \pm0.05$ and $c=4$, and material stiffnesses corresponding to yellow poplar LVL: 
$C_{(11)}=E_x/(1-\nu_{xy}\nu_{yx})=14.471$ GPa, $C_{(22)}=E_y/(1-\nu_{xy}\nu_{yx})=0.445$ GPa, $C_{(12)}=\nu_{xy}E_y/(1-\nu_{xy}\nu_{yx})=0.276$ GPa, $C_{(66)}=G_{xy}=0.247$ GPa, using values 
$E_x = 14.3$ GPa, $E_y = 0.44$ GPa, $G_{xy}= 0.247$ GPa, $\nu_{xy}=0.620$, $\nu_{yx}=0.019$, 
as reported in \cite{Janowiak01}.
The toy models were constructed by cutting elliptical slits ($0.05$ inch wide in the center) lengthwise on a $8$x$4$ inch sheet of $0.005$ inch thick polyester to which $0.0625$ inch diameter acrylic stiffening rods were glued lengthwise.  The sheets were contracted or expanded using tape across the slits.  A quantitative comparison of these models and the energy minima is not intended, but the qualitative correspondence is clear.
Despite swelling and shrinking having the same amplitude and slope, there is a small difference in the values of the parameter $\eta$ for the two minima. This difference seems to depend on the finite panel size, but may also be related to the use of the current area form in the energy.
The central regions of the shrunken and swelled shapes are saddle-like and dome-like, respectively, but each surface contains regions of both positive and negative Gaussian curvature, and with high anisotropy they are best qualitatively described as nearly cylindrical.
However, as with isotropic materials, perfect cylinders are not possible as they can always lower their energy through an isometric unbending into a piece of the plane.
Furthermore, as we will see below, other non-axisymmetric minima may arise at high anisotropy.

\begin{figure}[h!]
	\centering
	\includegraphics[width=0.75\linewidth]{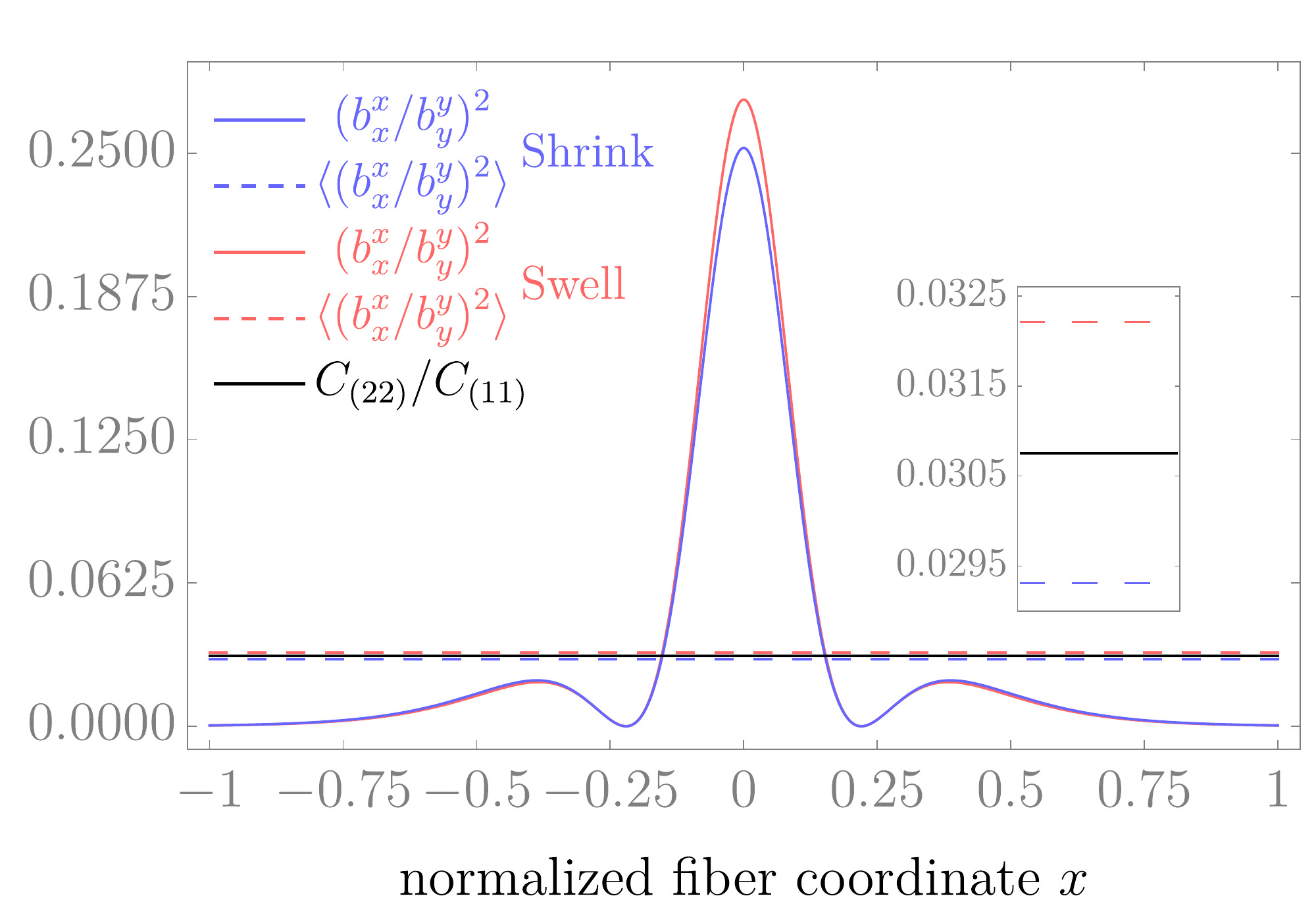}
	\caption{Ratios of squared curvatures $\left( b_x^x / b_y^y \right)^2$ and their area-weighted averages $ \langle \left( b_x^x / b_y^y \right)^2 \rangle \equiv \int \! \left( b_x^x / b_y^y \right)^2 U dx / \int U dx $ for the axisymmetric minima of Figure \ref{models}, compared with the naive prediction $ C_{(22)} / C_{(11)} $.}
	\label{RatioPlot}
\end{figure}

For such axisymmetric shapes, we can exploit the alignment of the principal curvatures with the principal material directions to predict the partition of bending energy in the structure.  The off-diagonal curvatures vanish, and a naive local minimization of the remaining terms $C_{(11)}(b_x^x)^2 + C_{(22)}(b_y^y)^2$ subject to a constraint on Gaussian curvature $K=b_x^xb_y^y$ suggests a preferred ratio of 
 $\left(b_x^x / b_y^y\right)^2=C_{(22)} / C_{(11)}$.   
 Since K changes sign on the surface, it is impossible for this relationship to hold everywhere.  However, it can be seen in Figure \ref{RatioPlot} that an area-weighted average of the ratio of squared curvatures falls very close to this value.
 Again we note a small asymmetry between the swelling and shrinking cases.

\begin{figure}[h!]
	\includegraphics[width=0.6\linewidth]{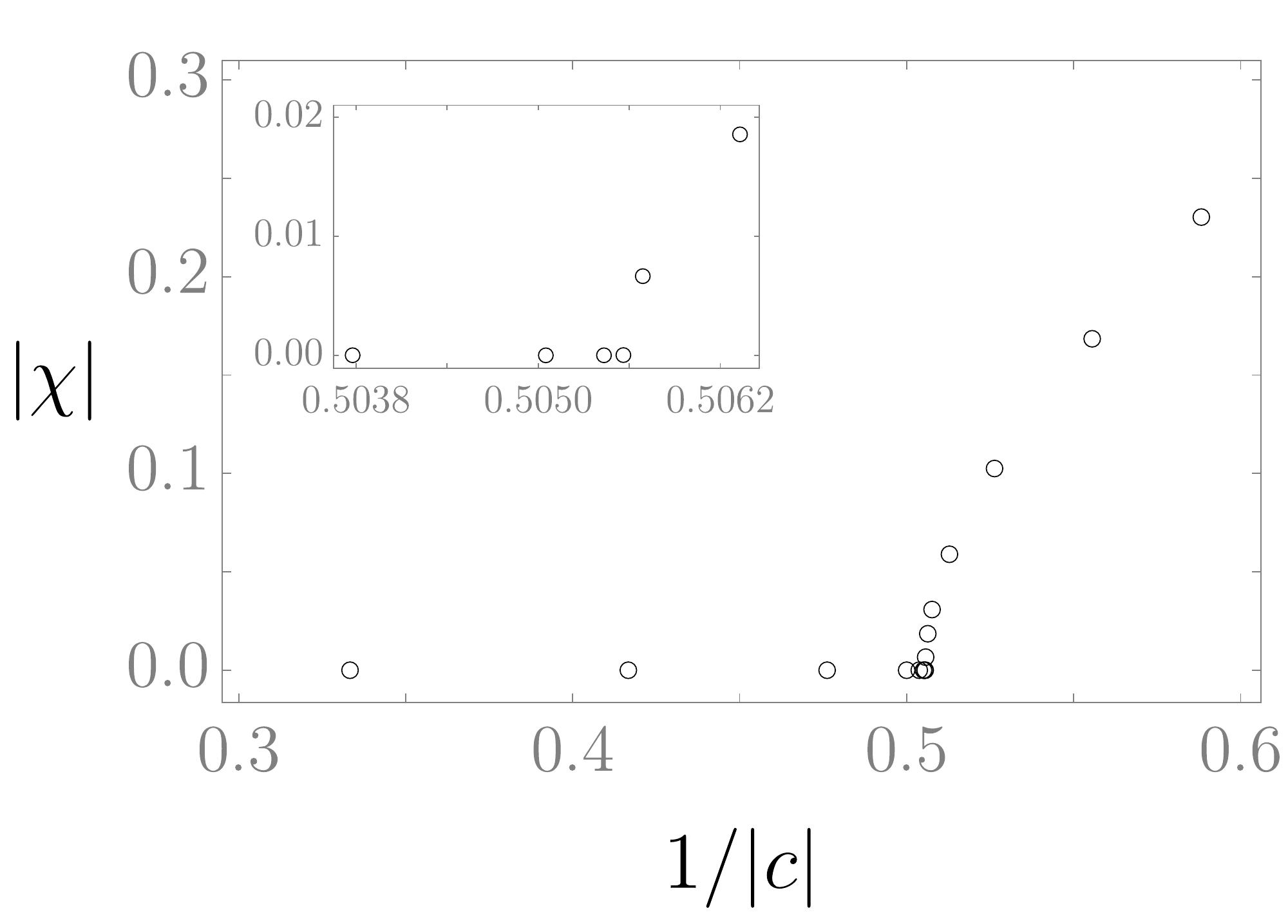}\\
	\includegraphics[width=0.6\linewidth]{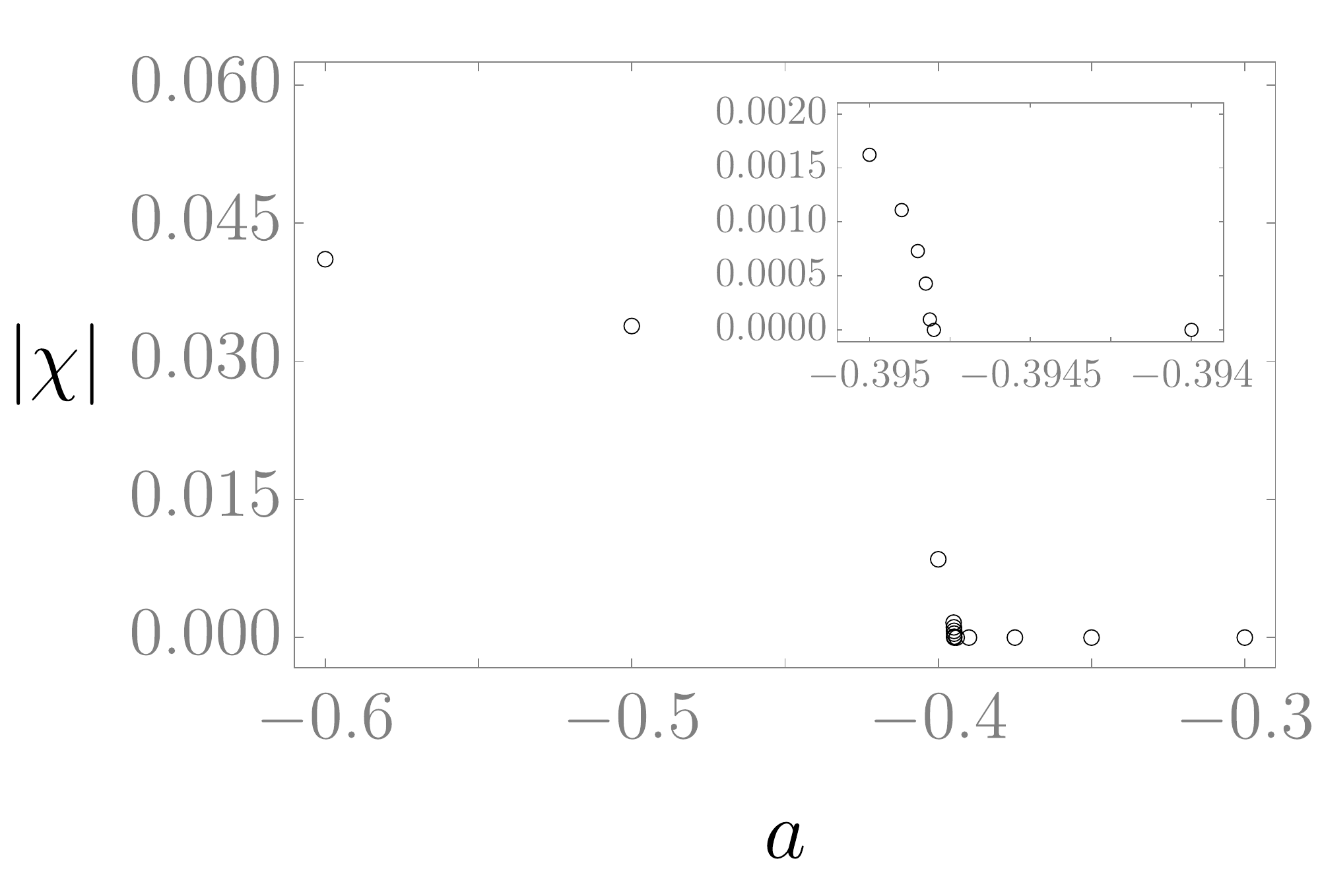}\\
	\includegraphics[width=0.6\linewidth]{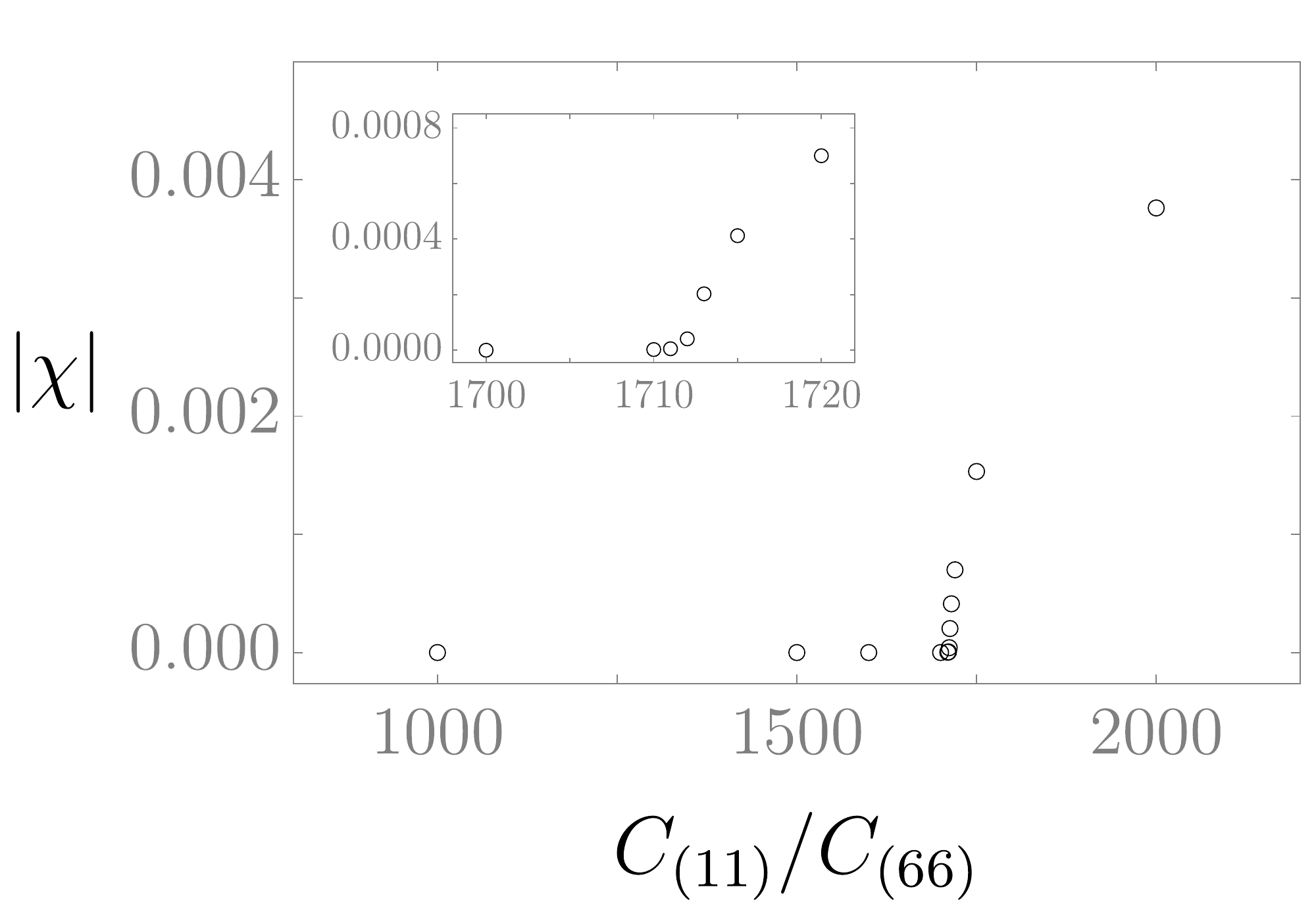}
	\caption{Values of $| \chi |$ for minimum energy embeddings of the metric with transverse swelling $g_{yy} = U^2 = [1+a\sech (cx)]^2$.  Top: varying the slope of the swelling $c$, with amplitude $a=-0.05$ and material property ratios $C_{(11)}/C_{(22)}=32.5$, $C_{(11)}/C_{(12)}=52.4$, and $C_{(11)}/C_{(66)}=58.6$.  Middle: varying $a$, with $c=4$ and the same material properties.  Bottom: varying $C_{(11)}$, with $a=-0.05$, $c=4$, and $\frac{1}{2}C_{(22)} = C_{(12)} = C_{(66)} =1$.	}
	\label{allplots}
\end{figure}

As swelling and material properties are varied, we observe pitchfork bifurcations to twisted ($\chi \ne 0$) minima for surfaces with central shrinking, several of which are presented in Figure \ref{allplots}.
Within the framework of the chosen function, we observe that either larger amplitude or greater delocalization of the shrinking favors twist.  
Twist can also be triggered by increasing the anisotropy.  This is intuitively reasonable, as twist is a means for the stiff fiber directions to reorient themselves more closely with the asymptotic directions along which curvature vanishes in a surface with negative $K$.
Although not shown in the figure, the post-bifurcation behavior can be non-monotonic at high values of anisotropy, indicating additional complexities not addressed here.

The three smaller bending energy terms in \eqref{kenmotsuenergy} are either clearly increasing or decreasing with increasing $\chi^2$, independently of the choice of swelling function $U > 0$.  However, the dominant $C_{(11)}$ term is not so straightforward.  An expansion in the twist parameter yields
\begin{align}
	U(b_x^x)^2 =\frac{1}{1-\eta^2(U')^2} \left( \eta^2U(U'')^2+ \frac{\eta^2U(U'')^2-2U''\left[1-\eta^2(U')^2\right]}{\eta^2U^2\left[1-\eta^2(U')^2\right]} \chi^2 \right) + O(\chi^4) \, . \label{bxx}
\end{align}
As both $U$ and $1-\eta^2(U')^2$ are positive, the requirement for $\chi = 0$ to be an energy maximum for a local patch of material is
\begin{align}
	0 < U'' < \frac{2\left[1-\eta^2(U')^2\right]}{\eta^2U} \, , \label{condition}
\end{align}
but of course the global shape is governed by the integral over \eqref{bxx} rather than this simple relation alone.
Note also that this is not the only way for twist to be favored by the system.  In our initial helicoidal example, the $b_x^x$ term was eliminated and thus the smaller terms in the energy chose the twisted state.

\section{Concluding discussion}

Our preliminary and restricted treatment of a seemingly simple problem of unidirectional gradients in transverse swelling of anisotropic elastic plates reveals that this physical system is richer than one might expect.
Beyond the phenomena revealed here, variations in shape and orientation of the plate perimeter are likely to yield new effects relevant to structures composed of anisotropic soft matter.
Certain aspects of our treatment relate to an industrial inverse problem, namely the potential for diagnosis of swelling distributions from the warped shapes adopted by engineered wood panels.
In isotropic materials, central swelling or shrinkage, or the presence of through-thickness gradients, have fairly straightforward signatures.
However, revisiting this driving question for anisotropic sheets, our findings are somewhat discouraging, particularly the complexity of the dominant bending term \eqref{bxx} as well as the possibility of multiple routes to twisted states.
Central shrinkage may lead to twisted shapes, as are common in wood panel manufacture, but can also have effects similar to central swelling that are in turn only subtly different from a through-thickness gradient in sufficiently anisotropic materials.
Yet though industry may lament 
 this unexpected complexity, it suggests both research avenues and design tools for programmable materials and structures.

\section*{Acknowledgments}
Support was provided by the Wood-Based Composites Center, a National Science Foundation Industry/University Cooperative Research Center (award 1624536-IIP).
We also acknowledge very informative interactions and hospitality from several Center members.

\bibliographystyle{unsrt}


\end{document}